\documentclass[english]{article}
\usepackage{mathptmx}
\usepackage[T1]{fontenc}
\usepackage[utf8]{inputenc}
\usepackage[letterpaper]{geometry}
\usepackage{babel}
\usepackage{textcomp}
\usepackage{amssymb}
\usepackage{graphicx}
\usepackage[authoryear]{natbib}
\usepackage[unicode=true,pdfusetitle,
 bookmarks=true,bookmarksnumbered=false,bookmarksopen=false,
 breaklinks=true,pdfborder={0 0 0},pdfborderstyle={},backref=false,colorlinks=false]
 {hyperref}

\makeatletter

\newcommand*\LyXThinSpace{\,\hspace{0pt}}
\providecommand{\tabularnewline}{\\}

\newcommand{\lyxaddress}[1]{
	\par {\raggedright #1
	\vspace{1.4em}
	\noindent\par}
}

\makeatother

\begin{document}
\title{Reply to ``Comment on ``Advanced Testing of Low, Medium, and High
ECS CMIP6 GCM Simulations Versus ERA5-T2m'' by N. Scafetta (2022)''
by Schmidt, Jones, and Kennedy (2023)}
\author{Nicola Scafetta}
\maketitle

\lyxaddress{$^{1}$Department of Earth Sciences, Environment and Georesources,
University of Naples Federico II, Complesso Universitario di Monte
S. Angelo, via Cinthia, 21, 80126 Naples, Italy.\\
}
\begin{center}
Graphical Abstract\\
\par\end{center}

\includegraphics[width=1\textwidth]{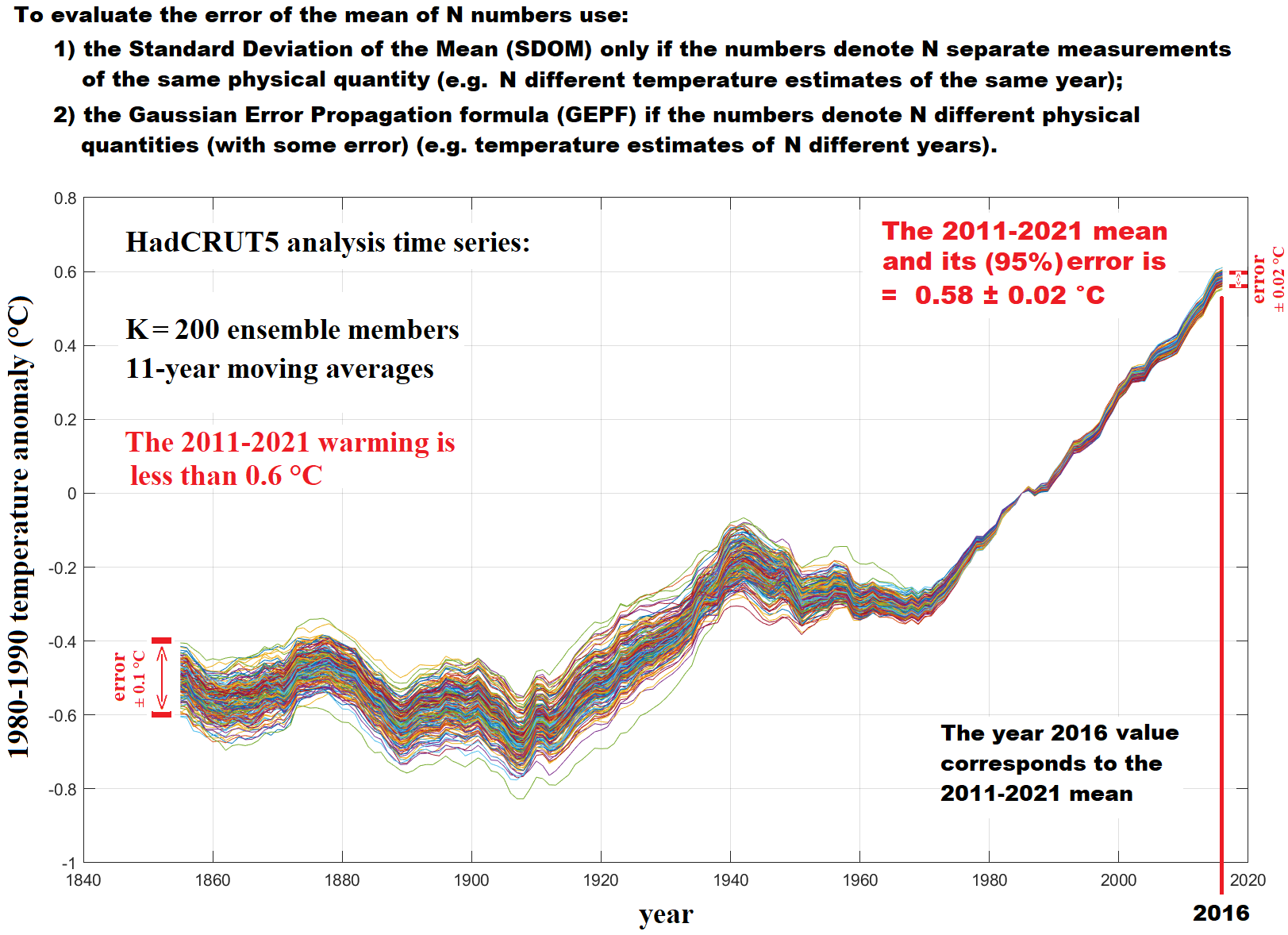}\\

\textbf{Cite as}: Scafetta N. (2023). Reply to ``Comment on ``Advanced
Testing of Low, Medium, and High ECS CMIP6 GCM Simulations Versus
ERA5-T2m'' by N. Scafetta (2022)'' by Schmidt, Jones, and Kennedy
(2023). \emph{Geophysical Research Letters} 50, e2023GL104960. https://doi.org/10.1029/2023GL104960

\newpage{}
\begin{abstract}
Schmidt, Jones \& Kennedy (SJK) \citeyearpar[https://doi.org/10.1029/2022GL102530, submited to GRL]{Schmidt}'s
critique of \citet[https://doi.org/10.1029/2022GL097716]{Scafetta}
is flawed. Their assessment of the error of the ERA-T2m 2011--2021
mean ($\approx0.10\;{^\circ}\mathrm{C}$) is 5-10 times overestimated
and contradicts published literature. SJK confused natural variability
with random noise, and mistook the error of the mean of a temperature
chronology for the stochastic error of the regression parameter $M$
of a nonphysical isothermal climate model ($T(t)=M$). SJK's allegations
regarding the internal variability of the models, the role of the
global climate model ensemble members, and other issues were partially
addressed in \citet[https://doi.org/10.1029/2022GL097716]{Scafetta}
and, later, more extensively in \citet[https://doi.org/10.1007/s00382-022-06493-w]{Scafetta(2023a)}
where \citet[https://doi.org/10.1029/2022GL097716]{Scafetta}'s conclusions
were confirmed.
\end{abstract}

\subsubsection*{Plain language summary}

Schmidt, Jones \& Kennedy (SJK) \citeyearpar[https://doi.org/10.1029/2022GL102530, submitted to GRL]{Schmidt}'s
assessment of the error of the ERA-T2m 2011--2021 mean ($\sigma_{\mu,95\%}=0.10\;{^\circ}\mathrm{C}$)
incorrectly assumes that, during such a period, the global surface
temperature was constant ($T(t)=M$) and that its interannual variability
($\Delta T_{i}=T_{i}-T(t_{i})=T_{i}-M$) was random noise. This is
a nonphysical interpretation of the climate system that inflates the
real error of the temperature mean by 5-10 times. In fact, the analysis
of the ensemble of the global surface temperature members yields a
decadal-scale error of about 0.01-0.02$\;{^\circ}\mathrm{C}$, as
reported in published records and deduced from the Gaussian error
propagation formula (GEPF) of a function of several variables (such
as the mean of a temperature sequence of 11 different years) \citep[chapters 3 and 9]{Taylor}.
Instead, SJK assessed such error using the standard deviation of the
mean (SDOM) \citep[chapter 4]{Taylor}, which is an equation that
can only be used when there exists a distribution of repeated measurements
of the same variable (which is not the present case). Furthermore,
SJK misinterpreted \citet[https://doi.org/10.1029/2022GL097716]{Scafetta}
and ignored published literature such as \citet[https://doi.org/10.1007/s00382-022-06493-w]{Scafetta(2023a)}
that already contradicted their main claim about the role of the internal
variability of the models and confirmed the results of \citet[https://doi.org/10.1029/2022GL097716]{Scafetta}.

\subsubsection*{Keywords: Global climate models; Global surface temperature; Chaotic
systems; Stochastic processes}

\subsubsection*{Keypoints}
\begin{itemize}
\item \citet[https://doi.org/10.1029/2022GL102530, submitted to GRL]{Schmidt}
assessed the error of the temperature mean by erroneously using the
SDOM instead of the error propagation formula.
\item The interannual temperature variability is physical signal, not noise;
the 2011--2021 ERA5-T2m mean error is not 0.10 °C, but 0.01-0.02
°C.
\item The role of the internal variability of the models was extensively
addressed in \citet[https://doi.org/10.1007/s00382-022-06493-w]{Scafetta(2023a)},
who confirmed \citet[https://doi.org/10.1029/2022GL097716]{Scafetta}.
\end{itemize}

\subsubsection*{Blog Comment}

Scafetta, N.: Comment and Reply to GRL on evaluation of CMIP6 simulations,
Climate Etc., (2023)\\
\href{https://judithcurry.com/2023/09/24/comment-and-reply-to-grl-on-evaluation-of-cmip6-simulations/}{https://judithcurry.com/2023/09/24/comment-and-reply-to-grl-on-evaluation-of-cmip6-simulations/}

\section{Introduction}

According to Schmidt, Jones \& Kennedy \citeyearpar{Schmidt} (SJK2023),
\citet{Scafetta} (S2022) contains ``\emph{numerous conceptual and
statistical errors that undermine all of the conclusions}''. SJK2023
reiterates the erroneous claims published in \emph{Real Climate} on
March 30, 2022 \citep{Schmidtetal2022a} and on October 10, 2022 \citep{Schmidt2022}.

Meanwhile, S2022 is now supported by the publication of other three
works: \citet{Scafetta(2023a),Scafetta(2023b)} provided a more comprehensive
analysis of the same topic; \citet{Lewis(2022)} reassessed downward
the likely range of the equilibrium climate sensitivity (ECS) previously
estimated by \citet{Sherwood}. These three studies support S2022's
conclusion that $\mathrm{ECS}\leq3.0\;{^\circ}\mathrm{C}$.

Herein, I demonstrate that SJK2023 is flawed. Section 2 overviews
S2022 and follow-up works and highlights some of SJK2023's key misinterpretations
of S2022. SJK2023's key criticism regarding the error of the mean---the
only part of their argument supported by a calculation---is thoroughly
rebutted in Section 3. Section 4 briefly addresses the internal variability
issue using the same results reported in S2022's Table 1; the extended
analysis is found in \citet{Scafetta(2023a)}. Section 5 briefly discusses
the other critiques.

\section{Overview of \citet{Scafetta} and Subsequent Works}

S2022 measured the global surface warming from Jan/1980--Dec/1990
to Jan/2011--Jun/2021 based on the ERA5--T2m record \citep{Hersbach}
and compared it with the ensemble average hindcasts produced by 38
CMIP6 global climate models (GCMs) downloaded from KNMI Climate Explorer
(\href{https://climexp.knmi.nl/start.cgi}{https://climexp.knmi.nl/start.cgi}).
The historical forcing functions (1850--2014) and three distinct
shared socioeconomic pathway scenarios (SSP2-4.5, SSP3-7.0, and SSP5-8.5;
2015--2100) were used for a total of 107 simulations \citep[Table 1]{Scafetta}.
S2022 aggregated the CMIP6 GCMs into three sub-ensembles, herein labeled
``macro-GCMs'', according to their ECS value: Low-ECS ($1.5<\mathrm{ECS}\leq3.0\;{^\circ}\mathrm{C}$);
Medium-ECS ($3.0<\mathrm{ECS}\leq4.5\;{^\circ}\mathrm{C}$), and High-ECS
($4.5<\mathrm{ECS}\leq6.0\;{^\circ}\mathrm{C}$). The Low-ECS macro-GCM
well hindcasted the observed warming; the Medium- and High-ECS macro-GCMs
consistently showed a warm bias, suggesting that these two macro-GCMs
may not be reliable climate predictors for climate-change policies.

SJK2023 did not acknowledge that S2022's goal was to test three macro-GCMs,
nor that S2022 examined three average simulations for each model (when
available), which partially accounted for the uncertainty related
to the GCMs' internal variability, nor that \citet{Scafetta(2023a)}
extensively analyzed all CMIP6 GCM simulations available on KNMI Climate
Explorer, which included 143 GCM ensemble average records and 688
GCM member simulations. Finally, \citet{Scafetta(2023b)} examined
the 175 CMIP6 GCM simulations that Schmidt proposed as indicative
of the CMIP6 GCMs \citep[Supplementary file]{Hausfather}. \citet{Scafetta(2023a),Scafetta(2023b)}
confirmed the warm bias of the Medium- and High-ECS macro-GCMs.

\section{The error of the mean}

\begin{figure}[!t]
\centering{}\includegraphics[width=0.95\textwidth]{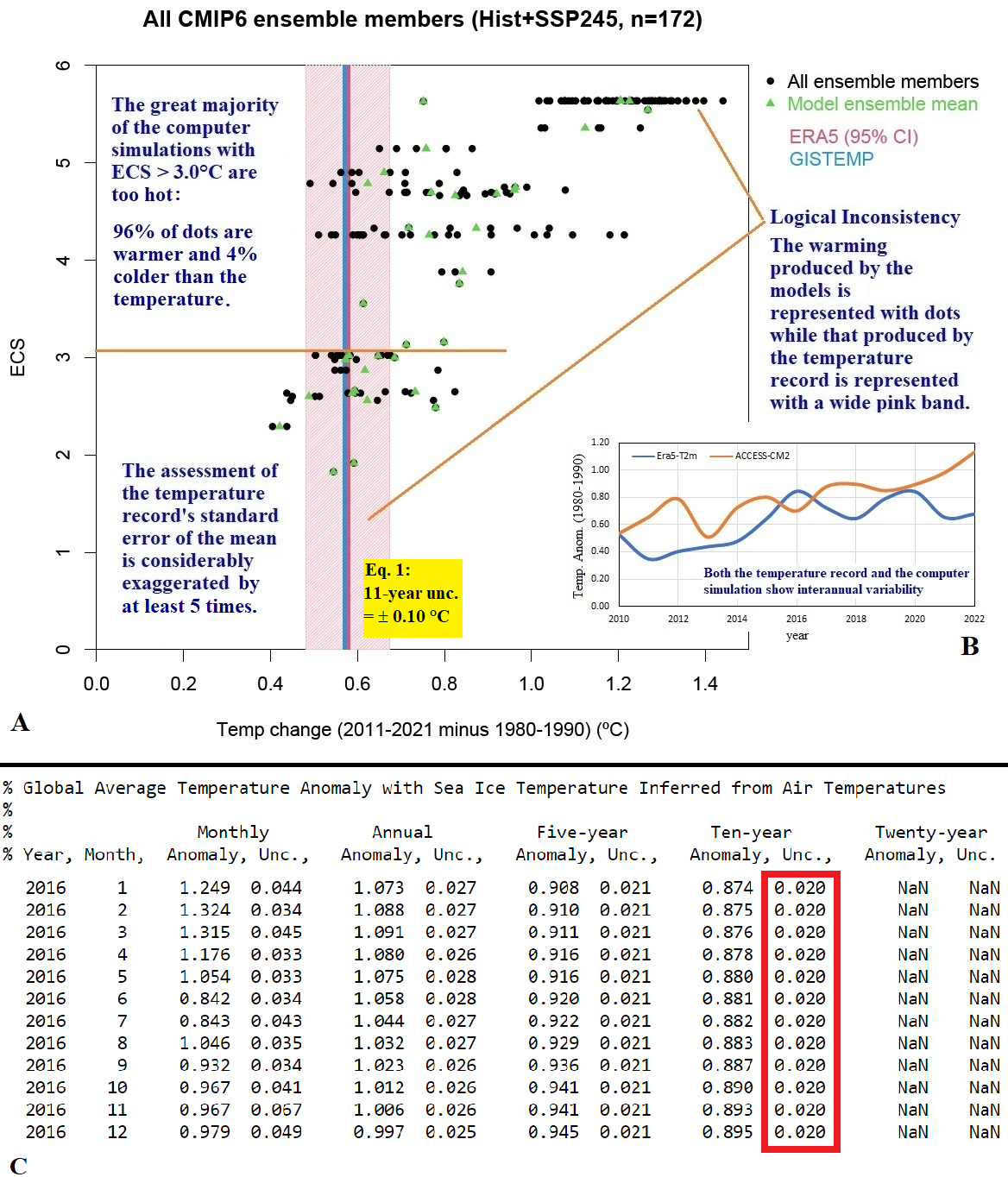}\caption{(A) SJK2023's figure displaying the actual and simulated temperature
changes from 1980--1990 to 2011--2021. The green dots represent
a small subset of the model ensemble averages examined in S2022. (B)
The insert shows that both the actual temperature record and computer
simulations present inter-annual variability. My critiques (highlighted
in the blue comments) are better detailed in the text. (C) Screenshot
of the Berkeley Earth's global surface temperature record (\protect\href{https://berkeley-earth-temperature.s3.us-west-1.amazonaws.com/Global/Land_and_Ocean_complete.txt}{https://berkeley-earth-temperature.s3.us-west-1.amazonaws.com/Global/Land\_and\_Ocean\_complete.txt})
with its estimated (95\% confidence) errors \citep{Rohde}. The red
box highlights the typical error of the 10-year mean centered in 2016
($\pm$0.020°C), which contradicts SJK2023's Eq. \ref{eq:1} ($\pm$0.10°C).}
\label{fig1}
\end{figure}

According to SJK2023, S2022 overlooked the ERA5-T2m error of the mean
from 2011 to 2021. SJK claimed that, in order to calculate the error
of the mean (95\% confidence), one must calculate the standard deviation
of the annual temperature anomalies around their 11-year mean and
multiply it by $1.96/\sqrt{N}$ (where $N=11$ is the number of years
from 2011 to 2021),

\begin{equation}
\sigma_{95\%}=1.96\times\frac{\sigma_{inter\,annual}}{\sqrt{N}}=1.96\times\frac{1}{\sqrt{N}}\sqrt{\frac{\sum_{i=1}^{N}(T_{i}-\mu)^{2}}{N-1}}=0.10\;{^\circ}C,\label{eq:1}
\end{equation}
where $T_{i}$ are the $N=11$ annual temperature values from 2011
to 2021 and

\begin{equation}
\mu=\frac{1}{N}\sum_{i=1}^{N}T_{i}\label{eq:2}
\end{equation}
is their mean over the 11-year period: see Figure \ref{fig1}A.

However, as \citet[Appendix]{Scafetta(2023a)} already explained,
Eq. \ref{eq:1} is wrong because the global surface temperature uncertainty
from 2011 to 2021 at the decadal mean is typically estimated to be
about $0.02\:{^\circ}\mathrm{C}$: see Figure \ref{fig1}C. In fact,
Eq. \ref{eq:2} is a function of $N$ separate quantities rather than
the mean of a distribution of $N$ repeated measurements of the same
quantity. Thus, as explained below, its error cannot be calculated
with the standard deviation of the mean (SDOM) of a (here-nonexistent)
distribution of one quantity \citep[chapter 4]{Taylor}, but must
be computed with the Gaussian error propagation formula (GEPF) of
a function of several quantities \citep[chapters 3 and 9]{Taylor}.

\subsection{Distinction Between the \textquotedblleft Mean\textquotedblright{}
and the  Regression Isothermal Model \textquotedblleft$T(t)=M$\textquotedblright}

GEPF establishes that the error of a generic function $q(z_{1},...,z_{N})$
of N different physical variables $z_{1},...,z_{N}$, each of which
is estimated with $K$ stochastic measurements $z_{i,k=1,...,K}$,
is

\begin{equation}
\sigma_{q}=\mathit{\sqrt{\sum_{i=1}^{N}\sum_{j=1}^{N}\frac{\partial q}{\partial z_{i}}\frac{\partial q}{\partial z_{j}}\cdot\sigma_{z_{i},z_{j}}}}=\mathit{\sqrt{\sum_{i=1}^{N}\left(\frac{\partial q}{\partial z_{i}}\right)^{2}\sigma_{z_{i}}^{2}+2\sum_{i=1}^{N-1}\sum_{j=i+1}^{N}\frac{\partial q}{\partial z_{i}}\frac{\partial q}{\partial z_{j}}\cdot\sigma_{z_{i},z_{j}}}}\label{eq:3}
\end{equation}
where

\begin{equation}
\sigma_{z_{i},z_{j}}=\mathrm{cov}(z_{i},z_{j})=\frac{1}{K-1}\sum_{k=1}^{K}(z_{i,k}-\bar{z}_{i})(z_{j,k}-\bar{z}_{j})\label{eq:4}
\end{equation}
is the covariance between $z_{i}$ and $z_{j}$: see \href{https://en.wikipedia.org/wiki/Propagation_of_uncertainty}{https://en.wikipedia.org/wiki/Propagation\_of\_uncertainty};
\citet[Eq. 13]{JCGM}; \citet[Eq. 9.9]{Taylor}. Note that $\sigma_{z_{i},z_{i}}=\sigma_{z_{i}}^{2}$
is the variance of each variable $z_{i}$, and the partial derivatives
are calculated at the average values $\bar{z}_{i}$. Eq. \ref{eq:3}
derives from a first-order Taylor series expansion of the function
$q(z_{1},...,z_{N})$ applied to stochastic variables.

\subsubsection{Definition of the Error of the Mean: \textquotedblleft$\sigma_{\mu,95\%}$\textquotedblright}

If $q(z_{1},...,z_{N})=\sum_{i=1}^{N}z_{i}/N=\mu$, we have $\partial q/\partial z_{i}=1/N$
and Eq. \ref{eq:3} becomes

\begin{equation}
\sigma_{\mu}=\frac{1}{N}\mathit{\sqrt{\sum_{i=1}^{N}\sigma_{z_{i}}^{2}+2\sum_{i=1}^{N-1}\sum_{j=i+1}^{N}\sigma_{z_{i},z_{j}}}}\label{eq:5}
\end{equation}
If the $N$ physical quantities $z_{i}$ are also characterized by
random and independent uncertainties, which means $\sigma_{z_{i},z_{j}}=0$
if $i\neq j$, Eq. \ref{eq:5} becomes

\begin{equation}
\sigma_{\mu}=\frac{1}{N}\mathit{\sqrt{\sum_{i=1}^{N}\sigma_{z_{i}}^{2}}}\label{eq:6}
\end{equation}
\citep[Eqs. 3.16 and 3.47]{Taylor}.

Therefore, if $T_{i}\pm\sigma_{i}$, for $i=1,2,...,N$, denote the
global surface temperatures of 11 different years and their errors
$\sigma_{i}$ are independent, their mean and its error are

\begin{equation}
\mu\pm\sigma_{\mu}=\frac{1}{N}\sum_{i=1}^{N}\left(T_{i}\pm\sigma_{i}\right)\approx\frac{1}{N}\sum_{i=1}^{N}T_{i}\pm\frac{1}{N}\sqrt{\sum_{i=1}^{N}\sigma_{i}^{2}}.\label{eq:7}
\end{equation}
Thus,

\begin{equation}
\sigma_{\mu,95\%}=1.96\times\frac{\sqrt{\sum_{i=1}^{N}\sigma_{i}^{2}}}{N}=1.96\times\frac{1}{\sqrt{N}}\sqrt{\frac{\sum_{i=1}^{N}\sigma_{i}^{2}}{N}}=1.96\times\frac{\overline{\sigma}}{\sqrt{N}},\label{eq:8}
\end{equation}
where

\begin{equation}
\overline{\sigma}^{2}=\frac{1}{N}\sum_{i=1}^{N}\sigma_{i}^{2}.\label{eq:9}
\end{equation}
Eq. \ref{eq:8} differs from Eq. \ref{eq:1} because $\overline{\sigma}^{2}$
is not defined as the variance of the data-set around its mean, but
as the mean of the variances of the data-points (Eq. \ref{eq:9}).

Eq. \ref{eq:5} should be used when the temperature record is represented
by an ensemble of K alternative records. The statistics resulting
from different temperature records is addressed in \citet{Scafetta(2023a)}.
In general, if the record permits to evaluate the data-point error
covariance, $\sigma_{\mu}$ must be calculated with Eq. \ref{eq:5},
which establishes that

\begin{equation}
0\leq\sigma_{\mu}\leq\frac{1}{N}\sum_{i=1}^{N}\sigma_{i},\label{eq:10}
\end{equation}
 \citep[Eqs. 3.17 and 3.48]{Taylor}.

For example, the mean between 10 and 20 is 15. If the numbers are
affected by independent Gaussian uncertainties, let's say $x=10\pm1$
and $y=20\pm1$, their mean is $15\pm0.7$ (Eq. \ref{eq:7}). If their
errors covary, the mean is $15\pm0$ if $\sigma_{x,y}=-1$ or $15\pm1$
if $\sigma_{x,y}=+1$ (Eq. \ref{eq:5}). However, Eq. \ref{eq:1}
calculates $15\pm5$ even when 10 and 20 indicate two different quantities
and are error-free, which is wrong.

\subsubsection{Definition of the error of the isothermal regression parameter $M$:
\textquotedblleft$\sigma_{M,95\%}$\textquotedblright}

In fact, Eq. \ref{eq:1} indicates something conceptually very different
from the error of the mean of $N$ different temperature quantities.
Eq. \ref{eq:1} indicates the standard deviation $\sigma_{M}$ of
the coefficient ``$M$'' of the regression model

\begin{equation}
T(t)=M,\label{eq:11}
\end{equation}
which postulates a strict physical temporal evolution of the global
surface temperature from 2011 to 2022.

The regression coefficient $M$ is numerically equal to $\mu$, but
their standard errors differ from each other because of the specific
physical meaning that Eq. \ref{eq:11} has. In fact, Eq. \ref{eq:11}
represents an isothermal climatic system. A model of this kind postulates
that from 2011 to 2021, the ``true'' global surface temperature
was constant and equal to $M$ for each year and, consequently, that
the temperature anomalies from $M$,

\begin{equation}
\Delta T_{i}=T_{i}-T(t_{i})=T_{i}-M,\qquad\qquad\mathrm{for}\qquad i=1,...,N,\label{eq:12}
\end{equation}
 were $N$ independent random numbers belonging to a zero-mean Gaussian
distribution with standard deviation equal to

\begin{equation}
\sigma_{\Delta T}=\sqrt{\frac{\sum_{i=1}^{N}(T_{i}-M)^{2}}{N-1}}.\label{eq:13}
\end{equation}
 Under such hypothesis, the error of $M$ is

\begin{equation}
\sigma_{M,95\%}=1.96\times\sigma_{M}=1.96\times\frac{\sigma_{\Delta T}}{\sqrt{N}}=1.96\times\frac{1}{\sqrt{N}}\sqrt{\frac{\sum_{i=1}^{N}(T_{i}-M)^{2}}{N-1}},\label{eq:14}
\end{equation}
which is Eq. \ref{eq:1}. If each data-point is also affected by a
measurement error $\sigma_{i}$, Eq. \ref{eq:14} is rewritten as

\begin{equation}
\sigma_{M,95\%}=1.96\times\sqrt{\frac{\sigma_{\Delta T}^{2}+\bar{\sigma}^{2}}{N}}>1.96\times\frac{\overline{\sigma}}{\sqrt{N}}=\sigma_{\mu,95\%},\label{eq:15}
\end{equation}
 where $\overline{\sigma}^{2}=\frac{1}{N}\sum_{i=1}^{N}\sigma_{i}^{2}$.

Eq. \ref{eq:15} shows that the error of the regression parameter
$M$ is different, and can also be significantly larger than the error
of the mean $\mu$ as deduced from the data uncertainties (Eq. \ref{eq:8}).
Thus, postulating Eq. \ref{eq:11} and that the anomalies $\Delta T_{i}$
are random noise artificially increases the variance of the temperature
data-points from $\sigma_{i}^{2}$ to $\sigma_{i}^{2}+\sigma_{\Delta T}^{2}$
and, therefore, also increases the uncertainty of their mean.

Eq. \ref{eq:13} can also be interpreted as a particular case of Eq.
\ref{eq:3} when $N=1$, $q(z)=z$ and $\partial q/\partial z=1$;
that is, when there only are $K$ repeated measurements of the ``same''
physical quantity $z$. Then, Eq. \ref{eq:14} (or Eq. \ref{eq:1})
is the SDOM of the distribution of the $K$ measurements of $z$ \citep[Eqs. 4.9 and 4.14]{Taylor}.

\subsection{Discussion}

The error of the 2011--2021 temperature mean is an observable that
solely depends on the given errors of the temperature data: if the
data are error-free, so is their mean. This error is unrelated to
the standard deviation of the constant coefficient of any hypothetical
regression model of the data (e.g. Eq. \ref{eq:11}), whose error
varies as the model changes.

SJK2023 assumed that the temperature anomalies $\Delta T_{i}$ (Eq.
\ref{eq:12}) were some kind of random noise produced by a mysterious
stochastic process that they labeled ``\emph{random nature}''. The
annual standard error of this unidentified stochastic process was
claimed to be $\sigma_{year,95\%}=0.1\sqrt{11}=0.33\;{^\circ}\mathrm{C}$,
which is one order of magnitude greater than the known annual stochastic
error of these data (cf. Table \ref{tab1}). Such an assumption physically
implies that from 2011 to 2021 the actual climate was isothermal,
$T(t)=M$, which is nonphysical. The same assumption statistically
implies that the 11 annual global surface temperatures from 2011 to
2021 are 11 repeated stochastic measurements of their 11-year mean,
which is incorrect.

Eq. \ref{eq:1} can only be used when dealing with repeated measurements
of the ``same'' physical quantity \citep[chapter 4]{Taylor}. This
occurs, for example, when several thermometers simultaneously measure
the temperature of the same location. In such instances, the measurement
discrepancies among the thermometers can be physically interpreted
as generated by stochastic processes. Yet, 11 numbers, each indicating
the annual global surface temperature of a different year, are not
11 repeated measurements of their 11-year mean; different years reflect
separate physical states of the climate, each with its own temperature,
and a 1-year period is not an 11-year period. Thus, the 11 temperatures
do not represent a stochastic distribution of one quantity but 11
different quantities. Therefore, the error of their mean cannot be
calculated with the SDOM (Eq. \ref{eq:1}), but only with the GEPF
(Eq. \ref{eq:6}, or, in general, Eq. \ref{eq:5}).

The interannual temperature variability is not ``noise'' or some
kind of ``error'' because it is due to well-known physical mechanisms
such as ENSO fluctuations, volcanic eruptions, variations in solar
activity, anthropogenic and natural warming trends, etc. Eq. \ref{eq:11}
can only be replaced with a realistic physical model of the type

\begin{equation}
T(t)=M+\Delta T_{physical}(t),\label{eq:16}
\end{equation}
where $M$ is the mean and the function $\Delta T_{physical}(t)$
captures the actual interannual physical component of the record,
which should coincide with the experimental measurements within their
errors of measure. $\Delta T_{physical}(t)$ can derive from external
forcings and/or internal mechanisms. \citet[Appendix]{Scafetta(2023a)}
showed that when this exercise is done, the error of $M$ approaches
the error of the mean (Eq. \ref{eq:8}). In fact, all temperature
records resemble the same temperature fluctuations, indicating that
those changes are primarily physical signal and not noise. As a result,
if no other kind of random error can be physically demonstrated, the
natural interannual variability of ERA5-T2m cannot increase the statistical
error of its 2011--2021 temperature mean.

\citet{Scafetta(2023a)} outlined other serious contradictions with
applying Eq. \ref{eq:1}. For example, by replacing the annual ($N=11$)
with the monthly ($N=132$) temperature record, the value of $\sigma_{\mu,95\%}$
calculated by Eq. \ref{eq:1} changes from about $0.108$ to $0.034\;{^\circ}\mathrm{C}$.
However, for records made of genuine stochastic Gaussian fluctuations,
$\sigma_{\mu,95\%}$ is independent of the time-scale of the data.
Thus, statistics excludes that ERA5-T2m consists of random Gaussian
fluctuations around a decadal mean. Therefore, SJK2023's calculation
is arbitrary because Eq. \ref{eq:1} highly depends on the temporal
resolution of the data. This inconsistency persists even if SJK2023's
isothermal model (Eq. \ref{eq:11}) were replaced with a linear model
(a referee proposal) because by simply interpolating the data with
additional points, the error of its constant coefficient converges
to zero  as N increases \citep[Eq. 8.16]{Taylor} because the data
variance around the model prediction will not change much since the
data-points are physically inter-correlated.

Figure \ref{fig1}A shows another severe logical, physical, and mathematical
inconsistency: SJK2023 estimated $\sigma_{M,95\%}$ (Eq. \ref{eq:1}
or \ref{eq:14}) just for the ERA-T2m temperature record and showed
it with a $\pm0.10\;{^\circ}\mathrm{C}$ pink-bar. However, for the
208 (36 green + 172 black dots) GCM hindcasts, they calculated $\sigma_{\mu,95\%}$
(Eq. \ref{eq:8}). The latter are correctly represented by dots since
the simulations are unaffected by any statistical error while showing
significant interannual variability like ERA-T2m, as Figure \ref{fig1}B
demonstrates. Yet, the same algorithm must be used for both the real
and synthetic temperature records.

It must also be mentioned that evaluating the 2011--2021 temperature
mean and its error does not depend on the GCMs' performance in modeling
the observed inter-annual climatic variability, nor on the possibility
of predicting the ``\emph{timing of El-Niño events}''. Moreover,
these events do not possess any ``\emph{random nature}'', as SJK2023
claimed by confusing a weakly chaotic system, like the climate, with
a stochastic process. Only the quantum world presents instances of
true randomness. In fact, while forecasting the ``\emph{timing of
El-Niño events}'' may be challenging (chaotic system), an El-Niño
event is a well-established fact when historically recorded. Yet,
Eq. \ref{eq:1} erroneously implies that the annual temperature data
are prone to stochastic errors that are so excessively large ($\sigma_{year,95\%}=0.33\;{^\circ}\mathrm{C}$)
that it may not even be possible to determine (stochastic process)
whether or not an El-Niño event has occurred in any given year.

In other words, since we are interested in determining what happened
from 2011 to 2021, the 2011--2021 ERA5-T2m interannual variability---which
represents the actual climatic chronology that occurred---cannot
be replaced by random data and, accordingly, Eq. \ref{eq:1} cannot
be applied to it. SJK somehow misunderstood that only the internal
variability of the GCMs can generate stochastic ensembles of interannual
hindcasts by randomly varying their initial conditions or internal
parameters. However, ERA5-T2m is not a GCM, but the given temperature
chronology that produces only one 2011--2021 temperature mean with
its own error of measure.

\subsection{Estimation of the likely \textquotedblleft error of the mean\textquotedblright{}
for the ERA-T2m 2011-2021 record}

S2022 did not report the error of the mean because the ERA-T2m monthly
or annual uncertainties had not been explicitly published \citep{Hersbach}.
However, such error was expected to be negligible because, as \citet{Scafetta(2023a)}
explained, over the last 50 years, the annual temperature uncertainties
are estimated to be 0.03-$0.05\;{^\circ}\mathrm{C}$ (95\% confidence)
by all global surface temperature records \citep{Lessen,Morice,Rohde}
and, if their error-covariance is ignored, such uncertainties should
be divided by $\sqrt{11}$ to get the error at the 11-year scale (Eq.
\ref{eq:8}).

For example, Table \ref{tab1} shows the HadCRUT5 annual mean values
from 2011 to 2021 together with their associated uncertainties. For
independent random errors, it is estimated that the error of the 11-year
mean is $\sigma_{\mu,95\%}\approx0.01\;{^\circ}\mathrm{C}$ (Eq. \ref{eq:8}),
as also online error-propagation-calculators confirm \citep{Bose,Lafarge,Wienand}.

\begin{table}[bh]
\centering{}%
\begin{tabular}{c|c|cccc}
\hline 
﻿ & ERA5-T2m & \multicolumn{4}{c}{HadCRUT 5.0.1.0 (infilled)}\tabularnewline
\hline 
﻿Time & Anomaly (${^\circ}C$) & Anomaly (${^\circ}C$) & Lower conf. (2.5\%) & Upper conf. (97.5\%) & $\sigma_{95\%,annual}$\tabularnewline
\hline 
2011 & 0.330 & 0.538 & 0.506 & 0.569 & 0.032\tabularnewline
2012 & 0.381 & 0.578 & 0.545 & 0.610 & 0.033\tabularnewline
2013 & 0.408 & 0.624 & 0.588 & 0.659 & 0.035\tabularnewline
2014 & 0.447 & 0.673 & 0.639 & 0.707 & 0.034\tabularnewline
2015 & 0.597 & 0.825 & 0.791 & 0.859 & 0.034\tabularnewline
2016 & 0.780 & 0.933 & 0.902 & 0.964 & 0.031\tabularnewline
2017 & 0.685 & 0.845 & 0.815 & 0.876 & 0.030\tabularnewline
2018 & 0.605 & 0.763 & 0.731 & 0.794 & 0.032\tabularnewline
2019 & 0.740 & 0.891 & 0.857 & 0.925 & 0.034\tabularnewline
2020 & 0.773 & 0.923 & 0.888 & 0.957 & 0.035\tabularnewline
2021 & 0.615 & 0.762 & 0.725 & 0.798 & 0.036\tabularnewline
\hline 
\multicolumn{1}{c}{mean =} & \multicolumn{1}{c}{0.578} & \multicolumn{1}{c|}{0.760} & (Eq. \ref{eq:9}) $\quad\quad\quad\;\;\;\;$ & $\quad\quad$ $\bar{\sigma}_{95\%,annual}$ = & 0.033\tabularnewline
\multicolumn{3}{c|}{} & \multicolumn{2}{c}{(Eq. \ref{eq:8}) $\quad\quad\sigma_{\mu,95\%}=\bar{\sigma}_{95\%,annual}/\sqrt{11}=$} & 0.010\tabularnewline
\hline 
\end{tabular}\caption{ERA5-T2m (1980--1990 anomalies). HadCRUT.5.0.1.0 (1961--1990 anomalies)
with their published statistical error (\protect\href{https://www.metoffice.gov.uk/hadobs/hadcrut5/data/current/download.html}{https://www.metoffice.gov.uk/hadobs/hadcrut5/data/current/download.html}).}
\label{tab1}
\end{table}

However, the decadal-mean temperature uncertainties were explicitly
published for the Berkeley Earth Land/Ocean Temperature record, which
reports $\sigma_{decadal.\mu,95\%}\approx0.02\;{^\circ}C$ since 1970
relative to the 1951--1980 period \citep{Rohde}. This publication
directly contradicts SJK2023's result derived from Eq. \ref{eq:1}.
Similarly, I processed the provided $K=200$ HadCRUT5 ensemble members
as 1980--1990 temperature anomalies with Eq. \ref{eq:5}, which evaluates
also the covariance among the individual members: the 2011--2021
mean was $0.58\;{^\circ}\mathrm{C}$ and $\sigma_{\mu,95\%}\approx0.02\;{^\circ}\mathrm{C}$
(by also including the coverage uncertainty): see Graphical Abstract.
This error-estimate is one-fifth of Eq. \ref{eq:1}, and is well-defined
because the same result is obtained with both the annual ($N=11$)
and monthly ($N=132$) records.

Thus, by analogy with the other published temperature records that
are nearly identical to ERA5-T2m, the ERA-T2m 11-year error was expected
to be $0.01\lessapprox\sigma_{\mu,95\%}\lessapprox0.02\;{^\circ}\mathrm{C}$,
which was negligible for S2022's analysis. In any case, Eq. \ref{eq:10}
establishes that $\sigma_{\mu_{11year},95\%}\leq\bar{\sigma}_{\mu_{1year},95\%}$,
and for the 2011--2021 period $\bar{\sigma}_{\mu_{1year},95\%}\ll0.10\;{^\circ}\mathrm{C}$
for all individual global surface temperature records \citep{Craigmile}.

\section{The Internal Variability of the Models}

According to SJK2023, S2022 ignored the internal variability of the
models and insisted that ``\emph{the full ensemble for each model
must be used}''.

As mentioned in Section 2, SJK2023's claim is incorrect because S2022
characterized each GCM using three independent average simulations.
That variability already provided an approximate estimate of the hindcast-range
related to each model's internal variability. Moreover, SJK2023 overlooked
that S2022's ultimate goal was to test three macro-GCMs where each
GCM ensemble average served as a sample of the ``internal variability''
of the respective macro-GCM.

\begin{figure}[!t]
\centering{}\includegraphics[width=1\textwidth]{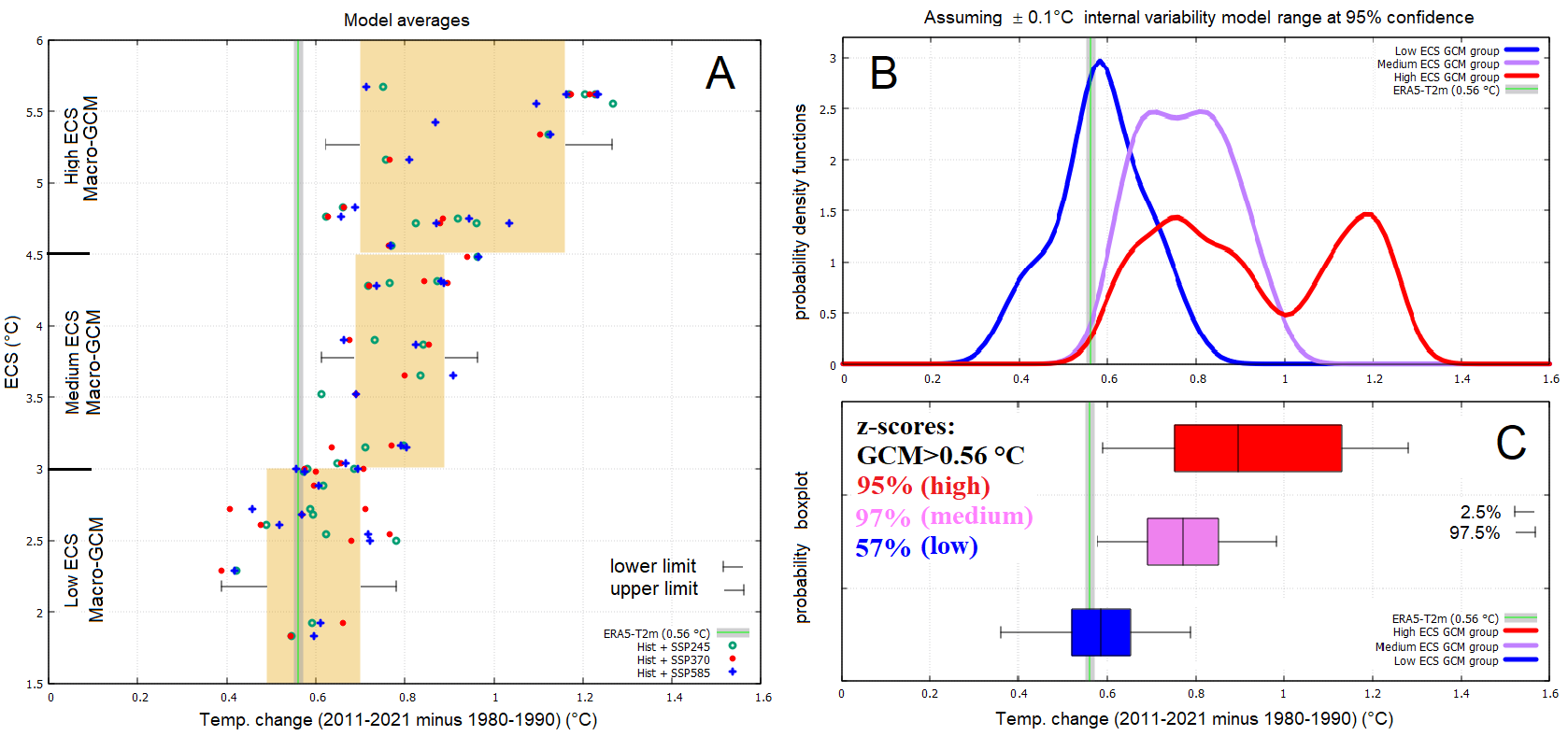}\caption{{[}A{]} Graphical reproduction of Table 1 in \citet{Scafetta} showing
three average hindcasts (blue, red and green dots) for each of 38
models. The yellow boxes represent the $\mu_{GCM}\pm\sigma_{GCM}$
range of the three macro-GCM distributions ($0.93\pm0.22$, $0.77\pm0.10$
and $0.58\pm0.10$ °C) with whiskers representing their extremes.
{[}B{]} Probability density functions (pdfs) for the Low-, Medium-
and High-ECS macro-GCMs assuming a $\sigma_{95\%}=0.10\;{^\circ}\mathrm{C}$
additional dispersion from each model average to simulate the GCM
internal variability. {[}C{]} Boxplots ($2.5\%,\;25\%,\;50\%,\;75\%,\;97.5\%$)
for the pdfs in B with their z-scores. The green vertical line represents
the estimated ERA5-T2m warming (Jan/2011--June/2021 mean) with a
$\pm0.01\;{^\circ}\mathrm{C}$ statistical error ($\pm\sigma$).}
\label{fig2}
\end{figure}

The test must determine whether the observed mean warming from 1980--1990
to 2011--2021 falls within the macro-GCM distribution of its hindcasts:

\begin{equation}
\mathrm{does}\;\mu_{obs}\pm\sigma_{obs}\in\mathrm{the\;macro\;GCM\;distribution\;of\;hindcasts}\;?\label{eq:17}
\end{equation}
Figure \ref{fig2}A graphically reproduces S2022's Table 1 with the
estimated actual error of the temperature record where, when possible,
each of the 38 models is represented by three independent simulations
(blue, red and green dots). All 70 average simulations of the 25 GCMs
of the Medium- and High-ECS macro-GCMs show a 1980--1990 to Jan/2011--Jun/2021
warming greater than $0.6\;{^\circ}\mathrm{C}$ and up to $1.3\;{^\circ}\mathrm{C}$,
whereas the ERA-T2m warming is $0.56\pm0.01\;{^\circ}\mathrm{C}$.

Later, \citet{Scafetta(2023a)} confirmed S2022's results by examining
all GCM simulations that were available (143 ensemble average and
688 member simulations), and he also provided a Monte Carlo methodology
to simulate the acceptable spread of the GCM hindcasts derived from
the GCM's internal variability, where 429,000 synthetic hindcasts
(about 9000 per model) were generated.

Actually, also SJK2023 confirmed S2022. For the Medium- and High-ECS
macro-GCMs, their figure (here Figure \ref{fig1}A) shows that, on
138 member simulations (black dots), 132 (96\%) are warmer and only
6 (4\%) are colder than the actual warming, which clearly reveals
the warm bias of these two macro-GCMs. Compare with the detailed analysis
of \citet[Table 2 and Figure 4]{Scafetta(2023a)}.

Figures \ref{fig2}B and \ref{fig2}C depict the statistical analysis
of the distribution of GCM hindcasts due to model internal variability
using the Monte Carlo methodology described in \citet{Scafetta(2023a)}
where each average simulation (three per every GCM, Figure \ref{fig2}A)
was assumed to present an additional stochastic dispersion given by
a Gaussian process with $\sigma_{95\%}=0.10\;{^\circ}\mathrm{C}$
(like the ERA5-T2m inter-annual variability calculated by SJK2023),
which was simulated by several hundred thousand random values. The
analysis uses the average simulation data from S2022's Table 1. Only
the Low-ECS macro-GCM is optimally compatible with the temperature
data within the 95\% confidence, whereas the Medium and High-ECS macro-GCMs
show a clear warm bias because the distribution of their hindcasts
is warmer than the data. In fact, the z-score percentiles of obtaining
GCM-hindcasts larger than $0.56\pm0.01\,\mathrm{{^\circ}C}$ are:
95\% (High-ECS-macro-GCM); 97\% (Medium-ECS-macro-GCM); 57\% (Low-ECS-macro-GCM).
Therefore, the Medium and High-ECS macro-GCMs perform rather poorly
in hindcasting the warming from 1980--1990 to 2011--2021 and should
not be used to formulate public policies.

The z-score is for $z=(\mu_{obs}-\mu_{GCM})/\sqrt{\sigma_{obs}^{2}+\sigma_{GCM}^{2}+0.05^{2}}$,
where $\mu_{obs}=0.56$ is the observed temperature mean, $\mu_{GCM}$
is the macro-GCM-hindcast-distribution mean, $\sigma_{obs}^{2}=0.01^{2}$
is the data variance, $\sigma_{GCM}^{2}$ is the macro-GCM variance
and $0.05^{2}$ is the assumed additional variance related to each
GCM internal variability.

Indeed, when a set of GCMs with comparable ECS properties is evaluated
as a macro-GCM, as done in S2022, even if a few models or some of
their simulations roughly hindcasted the observation (as some black
dots shown in Figure \ref{fig1} do) it would not demonstrate that
the macro-GCM performs well as a whole. In order to compare the actual
data with the macro-GCM hindcasts, statistical analyses like those
shown in Figure \ref{fig2} are required.

Here, I'm only concerned with the 11-year uncertainty of one record,
not the uncertainty resulting from variations among different temperature
records. In any case, all available global surface temperature records
show a 1980--1990 to 2011--2021 warming between 0.52 and $0.59\;{^\circ}\mathrm{C}$
\citep{Scafetta(2023a)}, which could also be significantly overestimated
according to alternative climatic records \citep{Scafetta(2023a),Scafetta(2023b),Spencer,Zou}.
The real 2011--2021 warming could be 0.4-0.6 °C, which is well-below
the Medium- and High-ECS macro-GCM distrubutions.

\citet{Scafetta(2023b)} proposed an ideal macro-GCM composed of 17
GCMs (from a collection of 41 GCMs) by selecting the top-performing
models based on their ECS and transient climate response (TCR) rankings.
Because of their low TCR, certain medium and high ECS models could
be added to the low-ECS category in this situation.

\section{The Areal Student t-test Analysis}

SJK2023 also critiqued the second analysis that S2022 proposed, which
was based on areal t-test statistics. The analysis was based on this
equations:

\begin{equation}
\Delta T_{j}=\frac{1}{N}\sum_{i=1}^{N}\Delta T_{i,j}^{m}-\Delta T_{j}^{o};\qquad t_{j}=\frac{\left|\Delta T_{j}\right|}{\sigma_{j}/\sqrt{N}},\label{eq:18}
\end{equation}
where $j=1,...,M$ are the grid cells, $i=1,...,N$ are the models,
$t_{j}$ represents the Student t-test variable and $\sigma_{j}$
is the standard deviation of the distribution of the GCM hindcasts:
$m$ indicates ``model'', and $o$ indicates ``observation''.
SJK2023 claimed that $\sqrt{N}$ should be replaced by 1.

SJK2023 did not propose a ``\emph{more appropriate test}'' but only
a complementary one that does not change the conclusion of S2022 that,
as the ECS lowers, the agreement between model hindcasts and actual
observations increases, on average, also at the local scale \citep[figure 5]{Scafetta}.
In fact, S2022's areal Student t-test maps can just be rescaled by
$1/\sqrt{N}$ if a reader is interested in studying the alternative
analysis. The test without $\sqrt{N}$ was adopted for the global
scale (Figure \ref{fig2}) where a dynamical comparison among the
various regions was not needed.

It is perplexing why SJK2023 asserted that the Student t-test would
not be adequate for evaluating the performance of the climate models
given that it is a well-known and reliable statistical instrument
of analysis. The major goal was to assess the local performance of
the three macro-GCMs. If $N$ increases, the rejection rate would
only rise if the models' reconstruction of the temperature changes
exhibited a systematic bias. In fact, for reliable models, the mean
divergence from the observed values should converge to zero ($\left|\Delta T_{j}\right|\rightarrow0$)
quicker than $1/\sqrt{N}$, and Eq. \ref{eq:18} will keep $t$ small
($t\rightarrow0$). Additionally, the (rather weak) dependence of
the t-test equation on $N$ can be easily solved by standardizing
the proposed test by using always a fixed number of models (e.g.,
$N=10$ or other), and, if necessary, averaging the areal t-test results
among these subsets.

In fact, the GCMs should hindcast the primary dynamical patterns produced
by the climate system, including those generated by the global atmospheric
and oceanic circulation. The fact that they cannot do it yet, does
not invalidate Eq. \ref{eq:18}. Simple energy balance models might
just replace the GCMs if they were not intended to reconstruct the
global circulation even at a timescale larger than the decadal one.
S2022's test successfully brought attention to the persistent problems
in climate dynamics that need to be solved in order to improve the
models.

On the other side, the ``\emph{more appropriate test}'' proposed
by SJK2023 would be inadequate for detecting specific dynamical biases
in the model ensemble because, for instance, one GCM simulation might
accurately reproduce the warming over China but not over Africa, whereas
another simulation might accurately reproduce the warming over Africa
but not over China (and so on). If the scaling provided by the factor
$\sqrt{N}$ in Eq. \ref{eq:18} is not used, and the two simulations
are just superimposed, the analysis would give the false impression
that the GCM reconstructs the temperatures in both China and Africa,
which would be misleading because none of the actual simulations would
do it.

S2022's recommended test is stringent, but statistical tests are generally
more useful when they reveal physical inconsistencies rather than
when they cover them under the umbrella of large stochastic error-bands.

\section{Conclusion}

SJK2023 made an incorrect use of statistics and physics because they
evaluated $\sigma_{\mu}$ of 11 different annual temperatures by using
the SDOM of a (here-nonexistent) distribution of one quantity instead
of the GEPF of a N-quantity function like the ``mean'', and confused
natural interannual chaotic variability with random noise.

Statistical equations only apply to the stochastic component of a
signal, not to its physical one, which is regulated by the laws of
physics. The uncertainty of any function of a temperature chronology
(such as its mean) arises only from the stochastic errors of the data-points
and their error-covariance (Eq. \ref{eq:5}). Whether the El-Niño
fluctuations are caused by internal mechanisms or external forcings,
or how well they can be predicted by modern GCMs, is irrelevant for
calculating the data errors at the decadal or any other timescale.
The observational errors must not be confused with the errors of the
regression parameters of a model-interpretation of the data.

More specifically, SJK2023 inflated the error of the ERA5-T2m 2011--2021
mean by 5--10 times by confusing it with the error of the regression
parameter $M$ of an isothermal climate model that postulates that
the interannual temperature variability is random noise, which is
nonphysical. Their $\sigma_{\mu,95\%}\approx0.10\,{^\circ}\mathrm{C}$
11-year error-estimate is contradicted by published temperature data
(e.g. Figure \ref{fig1}C) and was also arbitrarily calculated using
annual average values, as opposed to monthly ones, which produce with
Eq. \ref{eq:1} a very different result, $\sigma_{\mu,95\%}\approx0.03\,{^\circ}\mathrm{C}$.

Additionally, SJK2023 ignored that S2022's main result was confirmed
by three different sets (one for each SSP) of GCM average simulations
and that S2022's primary goal was to test three macro-GCMs characterized
by different ECS ranges. Finally, \citet{Scafetta(2023a)} answered
all of SJK2023's questions about the hindcast uncertainty of the models
and confirmed that the most likely ECS should be lower than 3 °C,
as Figure \ref{fig2}, which complements S2022, shows once again.

\newpage

\begin{figure}	
	\centering{}\includegraphics[width=1\textwidth]{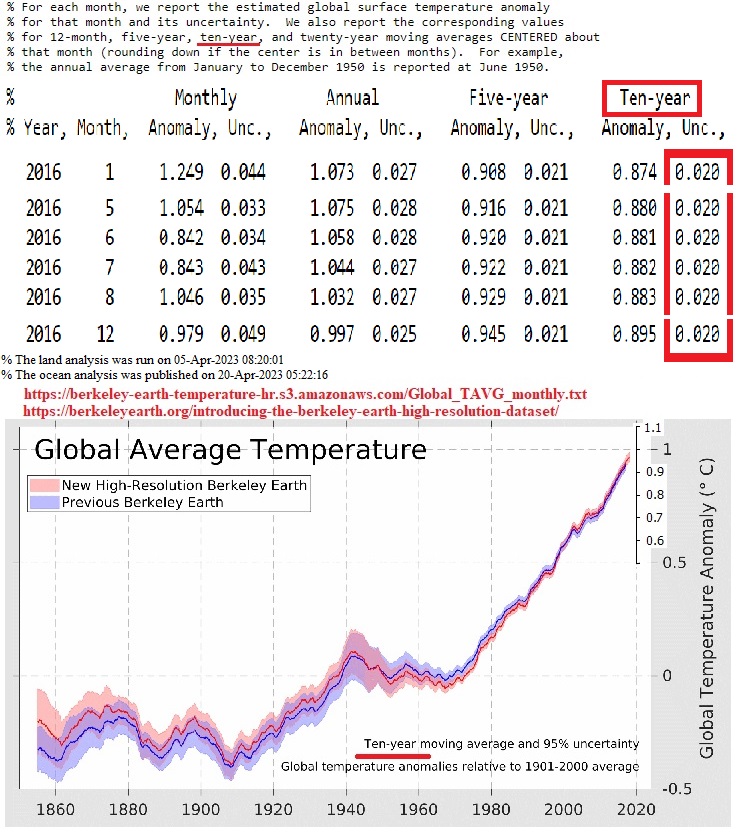}\caption{Berkeley Earth global average temperature, decadal scale with its
		95\% uncertainty range, which after 1980 is about $\pm0.02$ °C. }
\end{figure}

\end{document}